\pdfoutput=1
\documentclass[11pt,twoside,a4paper,cmspaper,final,collab]{cms-tdr}
\def\svnVersion{25c3326}\def\svnDate{2023/10/02}\def\cmsCernNoTag{CERN-EP-2022-151}\def\cmsCernDate{\today}\def\cmsMessage{Published in Physical Review Letters as \href{http://dx.doi.org/10.1103/PhysRevLett.131.101801}{\doi{10.1103/PhysRevLett.131.101801}.}}
\begin{document}\cmsNoteHeader{HIG-21-016}

\mathchardef\mhyph="2D
\renewcommand{\Pa}{\ensuremath{\mathcal{A}}\xspace}
\newcommand{\Pepem}{\Pep\Pem}
\newcommand{\egg}{\eta\to\gamma\gamma}
\newcommand{\agg}{\Pa\to\gamma\gamma}
\newcommand{\hgg}{\PH\to\gamma\gamma}
\newcommand{\haa}{\PH\to \Pa\Pa}
\newcommand{\hag}{\PH\to \Pa\Pa \to 4\gamma}
\newcommand{\gjet}{\gamma+\mathrm{jet}}
\newcommand{\mGG}{m_{\Gamma\Gamma}}
\newcommand{\Ea}{E_{\Pa}}
\newcommand{\ma}{m_{\Pa}}
\newcommand{\mG}{m_{\Gamma}}
\newcommand{\mao}{m_{\Gamma_1}}
\newcommand{\mat}{m_{\Gamma_2}}
\newcommand{\pol}{\mathrm{pol}(\mao,\mat)}
\newcommand{\mma}{2D-\ensuremath{\mG\,}}
\newcommand{\br}{\mathcal{B}}
\newcommand{\masb}{\ma\mhyph\mathrm{SB}}
\newcommand{\masr}{\ma\mhyph\mathrm{SR}}
\newcommand{\mh}{\ensuremath{m_{\PH}}\xspace}
\newcommand{\mhsb}{\mh\mhyph\mathrm{SB}}
\newcommand{\mhsblo}{\mh\mhyph\mathrm{SB_{low}}}
\newcommand{\mhsbhi}{\mh\mhyph\mathrm{SB_{high}}}
\newcommand{\mhsr}{\mh\mhyph\mathrm{SR}}
\newcommand{\mhsrmasb}{\masb\,\cap\,\mhsr}
\newcommand{\mhsrmasr}{\masr\,\cap\,\mhsr}
\newcommand{\mhsbmasb}{\masb\,\cap\,\mhsb}
\newcommand{\mhsbmasr}{\masr\,\cap\,\mhsb}
\newcommand{\zee}{\PZ \to \Pepem}
\ifthenelse{\boolean{cms@external}}{\renewcommand{\CL}{C.L.\xspace}}{}
\newcommand{\boost}{\gamma_{\mathrm{L}}}
\newcommand{\ctau}{c\tau_0}
\newcommand{\tbt}{$3{\times}3$ }
\newcommand{\mol}{Moli\`ere }
\newcommand{\rmol}{r_{\mathrm{M}}}
\newcommand{\chgiso}{\mathcal{I}_{\text{ch}}}
\newcommand{\muhat}{\hat{\mu}}

\cmsNoteHeader{HIG-21-016}
\title{Search for exotic Higgs boson decays \texorpdfstring{$\PH \to \Pa\Pa \to 4\gamma$}{H to AA to 4 photons} with events containing two merged diphotons in proton-proton collisions at \texorpdfstring{$\sqrt{s} = 13\TeV$}{sqrt(s) = 13 TeV}}

\date{\today}

\abstract{ We present the first direct search for exotic Higgs boson decays $\haa$, $\agg$ in events with two photonlike objects. The hypothetical particle $\Pa$ is a low-mass spin-0 particle decaying promptly to a merged diphoton reconstructed as a single photonlike object. We analyze the data collected by the CMS experiment at $\sqrt{s} = 13\TeV$ corresponding to an integrated luminosity of $136\fbinv$. No excess above the estimated background is found. We set upper limits on the branching fraction $\br(\PH \to \Pa\Pa \to 4\gamma)$ of (0.9--3.3)$ \times 10^{-3}$ at 95\% confidence level for masses of $\Pa$ in the range 0.1--1.2\GeV.} 

\hypersetup{
pdfauthor={CMS Collaboration},
pdftitle={Search for exotic Higgs boson decays H to AA to 4 photons with events containing two merged diphotons in proton-proton collisions at sqrt(s) = 13 TeV},
pdfsubject={CMS},
pdfkeywords={CMS, exotic Higgs decays, boosted Higgs, merged photons, machine learning}
}

\maketitle

Despite current constraints on the Higgs boson coupling to standard model (SM) particles from experiments at the CERN LHC~\cite{hcoupling,atlascoupling}, the Higgs sector remains an important area for physics searches beyond the SM (BSM). This is possible because the Higgs sector can potentially access BSM states that do not directly participate in the SM gauge interactions, called SM neutral. The simplest extensions of the Higgs sector include exotic decays of the form $\haa$, where $\PH$ is the 125\GeV boson with a non-SM decay, and $\Pa$ is a hypothetical new spin-0 particle decaying to a pair of SM particles~\cite{hexo}. Such decays are found in BSM models containing an additional SM-neutral singlet~\cite{hexo,hexo2} and are of interest in searches for axionlike particle (ALP) production~\cite{pqtheory,axioncollider,axionreview}, including analyses performed at CMS~\cite{cmshaa2m2t,cmshaa4m,cmshaa2m2tlow,cmshaa2m2b,cmshaa2m2tboost}. In ALP models, $\Pa$ is identified as a $CP$-odd spin-0 particle where $CP$ is the charge-conjugation and parity operator. The experimental search presented in this Letter is insensitive to the $CP$ quantum numbers of $\Pa$, since its polarization is not measured.

For particle $\Pa$ masses $\ma \lesssim 1\GeV$, decays to many of the heavier fermions become inaccessible and the diphoton decay mode $\agg$ becomes particularly relevant~\cite{bogdanhaa,hexo}. Generic $\agg$ searches are prevalent in astrophysics, cosmology~\cite{axionsearch1,axionsearch2,pdgreview2020}, and particle collider experiments~\cite{gluexagg,axionlead} because of the potential impact of $\Pa$ on stellar formation and the evolution of the early Universe. The particle $\Pa$ is also a potential dark matter candidate~\cite{axionastro,axioncosmo,axioncosmo2,axiondmreview}.

A measurement of the diphoton invariant mass spectrum at the LHC in this $\Pa$ mass regime is challenging if $\Pa$ is produced from the decay $\haa$. In the $\Pa$ mass range of $\ma < 0.4\GeV$, $\Pa$ will have a Lorentz boost $\boost \gtrsim 150$, where $\boost = \Ea/\ma$ and $\Ea$ is the $\Pa$ energy. In the CMS experiment, this boost corresponds to a distance between the photons from the $\agg$ decay at the electromagnetic calorimeter (ECAL) that is equal to or less than the \mol radius ($\rmol$) of the ECAL material. The two photons will be predominantly reconstructed as one photonlike object (labeled $\Gamma$) by the standard CMS photon reconstruction algorithm, as described in Ref.~\cite{e2e-massreg}. If $\Pa$ decays promptly, the $\hag$ decay will lead to a measured two-photon invariant mass $\mGG$ peak degenerate with that of the SM $\hgg$ decay. The $\hag$ signal will then be hidden by reconstructed SM $\hgg$ events~\cite{hgg,hcoupling}. With $\Pa$ decays to more-massive particles being inaccessible, $\Pa$ can also be long-lived~\cite{hexo}.

To separate a possible signal from the SM background, we present the
first search that directly measures the invariant mass spectrum of
merged diphoton candidates $\Gamma$ reconstructed in events resembling
an SM $\hgg$ final state. We assume that each $\Pa$ in the $\haa$ decay
has the same mass and decays promptly to photons, though we provide
estimated results where these assumptions are relaxed. The direct search
is made possible by a novel particle reconstruction technique that is
able to measure the invariant mass, $\mG$, of merged diphoton
candidates, something not possible with standard CMS reconstruction
software. The technique utilizes deep learning algorithms trained
directly on ECAL energy deposits to estimate $\ma$ in a so-called
end-to-end $\mG$ reconstruction~\cite{e2e-massreg}. It is applied for
the first time at CMS to probe masses from 0.1\GeV (corresponding to
$\boost \approx 600$ with the diphotons collimated enough to enter the
same ECAL crystal), to 1.2\GeV (corresponding to $\boost \approx 50$ and
diphotons separated by around 3$\rmol$ at the ECAL). Tabulated results
for this analysis can be found in~\cite{hepdata}. Signal events with larger $\ma$ have at least one of the $\agg$ decays reconstructed as two distinct photons and are not studied. We note that the technique requires access to the full CMS event content which is not readily accessible because of storage constraints. No special event triggers are otherwise required.

The analysis is performed using proton-proton ($\Pp\Pp$) collision data
at $\sqrt{s}=13\TeV$, collected by the CMS detector at the LHC from 2016
to 2018, corresponding to an integrated luminosity of $136\fbinv$. The CMS apparatus~\cite{cms} is a multipurpose, nearly hermetic detector, designed to trigger on~\cite{CMS:2020cmk,cmstrigger} and identify electrons, muons, photons, and hadrons~\cite{CMS:2015xaf,CMS:2018rym,CMS:2015myp,CMS:2014pgm,CMS:2020uim}. A global reconstruction algorithm~\cite{CMS:2017yfk} combines the information provided by the all-silicon inner tracker and by the lead tungstate ECAL and brass-scintillator hadron calorimeters, operating inside a 3.8\unit{T} superconducting solenoid, with data from gas-ionization muon detectors embedded in the solenoid return yoke, to build \PGt leptons, jets, missing transverse momentum, and other physics objects~\cite{CMS:2018jrd,CMS:2016lmd,CMS:2019ctu}. The integrated luminosities for the 2016--2018 data-taking years have 1.2--2.5\% individual uncertainties~\cite{CMS-LUM-17-003,CMS-PAS-LUM-17-004,CMS-PAS-LUM-18-002}, while the overall uncertainty for the 2016--2018 period is 1.6\%.

The $\agg$ decay is primarily detected in the ECAL, which contains a barrel section covering the pseudorapidity range~$\abs{\eta} < 1.48$ and two endcap sections on either end extending the range to $\abs{\eta} < 3$. The crystal $\rmol$ is comparable to the barrel crystal width of 2.2\cm, resulting in more than $90\%$ of the energy of photons converting only in the ECAL barrel to be laterally contained within a \tbt crystal matrix~\cite{CMS:2015myp}.

A sample of events containing two reconstructed photonlike objects ($\Gamma\Gamma$) is selected for analysis. We assume that each reconstructed photonlike object $\Gamma$ corresponds to a single merged diphoton candidate. Events selected for study are required to pass trigger, photon reconstruction, and kinematic selection requirements similar to those used in the CMS SM $\hgg$ analysis~\cite{hgg}. Key differences in the photon reconstruction criteria are described in the following paragraph. The SM background processes contributing to the selected sample are composed of non\-re\-so\-nant quantum chromodynamic (QCD) dijet and $\gjet$ production, prompt-diphoton production, and $\hgg$ decays.

Since the $\ma$ range we study spans a wide range of $\Pa$ boosts, the
opening angles of the $\agg$ decays vary as well~\cite{e2e-massreg}. We
thus apply a looser requirement on the output of the multivariate photon
tagger~\cite{hgg}, which then accepts more photons than for the CMS
$\hgg$ search. This increases the contribution from QCD background
processes containing hadronic jets, which may be reconstructed as
photonlike objects if they radiate an energetic photon or contain one or
more neutral-meson decays ($\Pgpz/\egg$). To compensate, we set a more
restrictive requirement on the transverse momenta of charged particles
in a cone around the selected photon, $\chgiso$~\cite{hgg}. Since
neutral mesons are mainly produced inside hadronic jets, they are more
likely to be accompanied by charged particle tracks. Thresholds for the
above requirements are chosen to maximize the significance of a possible
$\hag$ signal over the $\ma$ range of interest. After all selection
criteria are applied, signal events are selected with an estimated
efficiency of 8\%--24\%, decreasing with $\ma$, while background events
are reduced by $>$99\%, as determined from simulation.

To simplify the application of the new $\mG$ reconstruction technique, only events with both $\Gamma\Gamma$ reconstructed in the barrel section of the ECAL are analyzed. These account for about two thirds of the total expected signal yield. Events with more than two selected photons passing the selection criteria are not used to maintain orthogonality with a complementary CMS search at higher $\ma$ where the $\agg$ photons are fully resolved~\cite{h4gTanvi}.

To discriminate $\hag$ events from background, the end-to-end $\mG$ reconstruction technique is used to measure the two-dimensional (2D) invariant mass spectrum, $\mao$ versus $\mat$ (labeled \mma), where $\Gamma_{1(2)}$ is the higher- (lower-) energy reconstructed photon. Each $\Gamma$ is assumed to correspond to a single $\agg$ leg of the presumed $\hag$ decay. We construct signal (S) \mma templates for each signal mass hypothesis $\ma =$ 0.1--1.2\GeV, in 0.1\GeV steps, using simulation. This defines our signal model at each $\ma$ hypothesis. The relative $\mG$ resolution varies from approximately 100\% to 20\% for $\ma = 0.1$--1.2\GeV~\cite{e2e-massreg}. Background ($B$) \mma templates are constructed for the SM processes contributing to our selected sample, derived from data and simulation. The sum of these background templates defines our background model, representing the SM-only hypothesis. A scan over different signal $\ma$ hypotheses is performed. At each point, maximum likelihood estimation (MLE)~\cite{lhoodlhc} is used to extract the best fit between the observed \mma spectrum and the signal-plus-background model, $\mu S+B$, for some signal strength parameter $\mu$. The detection of a potential $\hag$ contribution at a given $\ma$ is expressed in terms of the significance of the extracted best fit signal strength.

Simulated $\hag$ events are generated with \MADGRAPH
\MCATNLO~\cite{mg5mcatnlo} at leading order using the SM + dark vector +
dark Higgs model~\cite{hexo,hexo2} and a Higgs boson mass of 125\GeV. A
rounded mass value of 125\GeV is chosen as this search is insensitive to
the exact value of the Higgs boson mass. The Higgs boson is produced
with up to one associated jet, and the $\Pa$ has a negligible
lifetime. The events are fully simulated in the CMS detector and account
for effects from the underlying event and additional $\Pp\Pp$ collisions
in the same or nearby bunch crossings. Signal samples are generated
separately for each $\ma$ hypothesis.

The simulated $\hag$ events are used to construct the signal \mma templates. Theoretical estimates of the $\haa$ production cross section vary depending on the model assumptions. Our final results, however, are expressed in terms of the signal branching fraction $\br(\hag)$, which is independent of the choice of cross section used to normalize the signal templates.

Background \mma templates are derived to account for the $\hgg$ contribution and all other non\-re\-so\-nant processes. The SM $\hgg$ \mma template is found from simulation. The $\hgg$ events are generated with \MADGRAPH \MCATNLO~\cite{mg5mcatnlo} at next-to-leading order, with up to two extra jets in the matrix element calculation using \textsc{FxFx} merging~\cite{fxfx}. The generated events are fully simulated in the CMS detector in the same way that the signal events described earlier are simulated. The normalization of the template is determined by fixing the template yield to the integrated luminosity of the data sample times the SM theoretical prediction for the total inclusive $\PH$ production cross section~\cite{higgswg} times the $\hgg$ branching fraction and efficiency to pass selection criteria.

\begin{figure*}[!htbp]
\centering
\includegraphics[height=.33\linewidth]{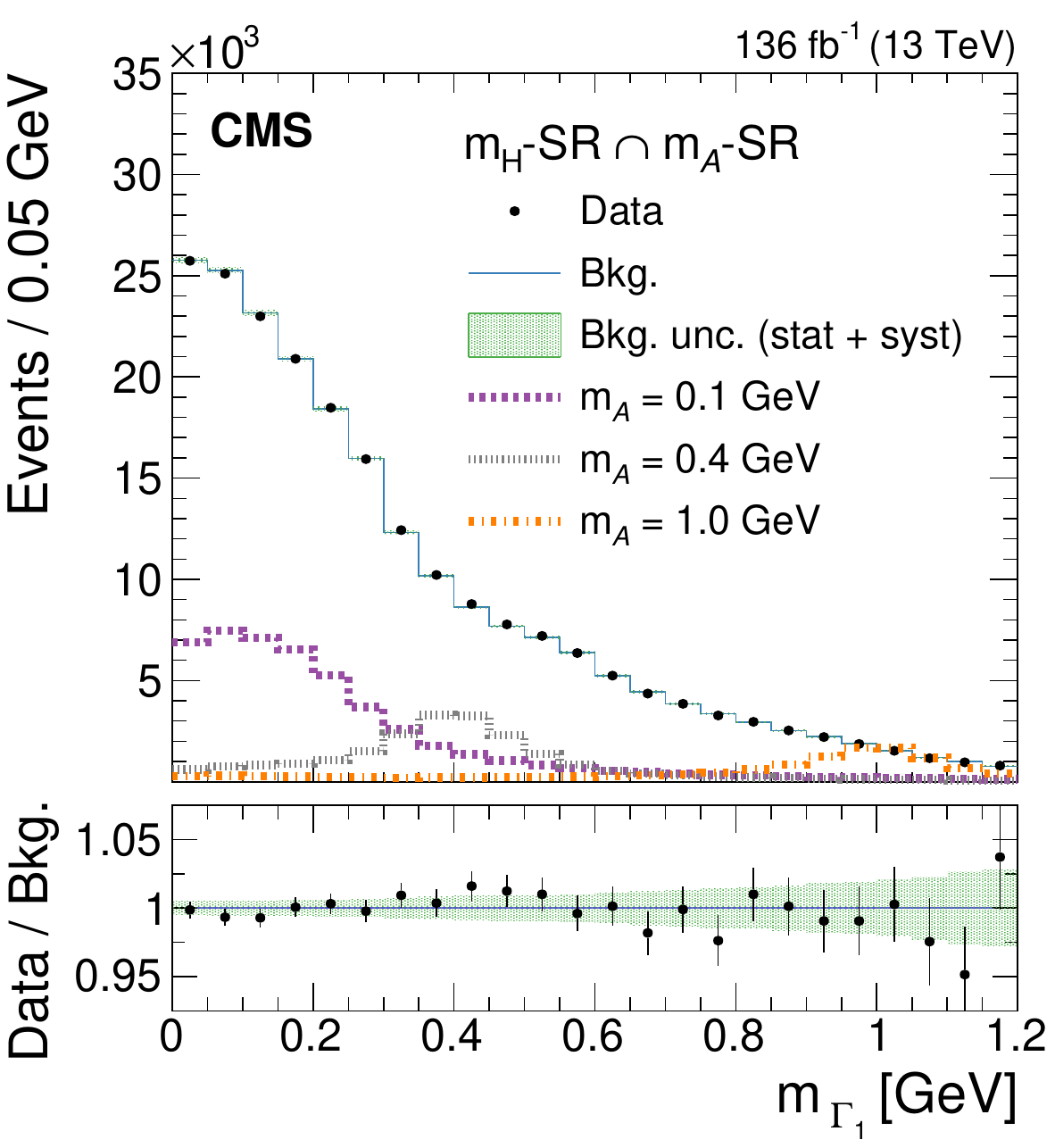}
\includegraphics[height=.3\linewidth]{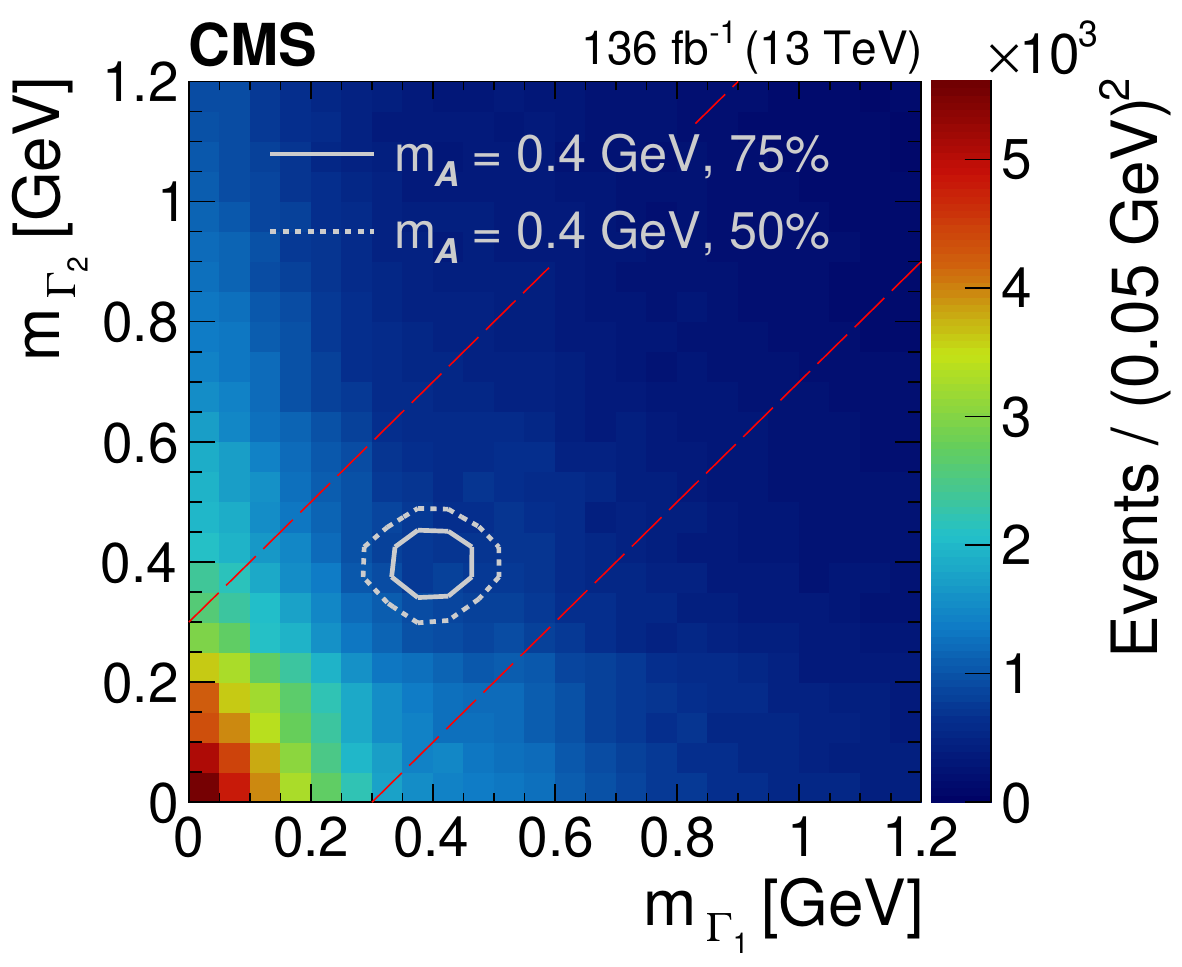}
\hspace{3pt}
\includegraphics[height=.33\linewidth]{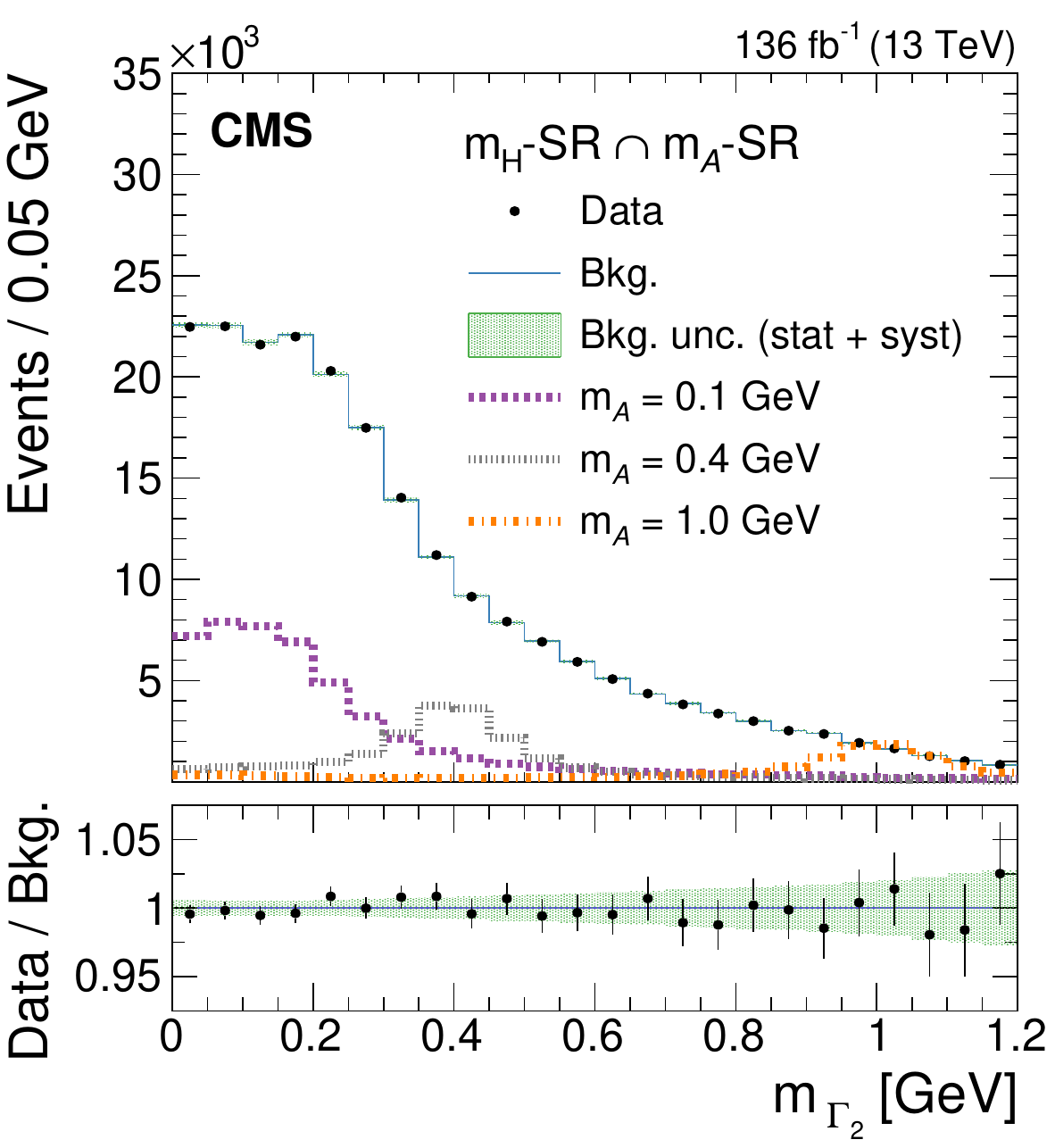}
\caption{Mass distributions from selected events in data. Center: the \mma distribution for data events in the $\mhsr$. The red dashed lines indicate the $\masr$ boundaries. The contours of simulated $\hag$ events for $\ma=0.4\GeV$ are plotted for $75\%$ (solid contour) and $50\%$ (dotted contour) of the distribution maximum. The corresponding $\mao$ (left) and $\mat$ (right) projections for the overlap of the $\mhsr$ and $\masr$ are also shown. The data distributions (black points) are plotted against the total predicted background distributions (blue curves). The statistical uncertainties in the former are negligible and the total uncertainties in the latter are barely visible as green bands. The spectra of simulated $\hag$ events for $\ma=0.1$ (purple dashed curve), $0.4$ (gray dotted curve), and $1.0\GeV$ (orange dash-dotted curve) are also provided. They are each normalized to the value of $\br(\hag)$ that is expected to be excluded by the background model (described under the \CLs criterion in our results) times $10^3$. The black points in the lower panels of the left and right plots give the ratios of the data to the predicted background distributions. The vertical bars represent the statistical uncertainties in the former, and the green bands represent the total uncertainty in the latter.}
\label{fig:2dma-bkg-obs}
\end{figure*}

An overall background \mma template for all other non\-re\-so\-nant processes is estimated from data. We divide the selected events in data into regions based on $\mGG$ and \mma. Since the two $\Pa$ particles come from the decay of a Higgs boson with a mass of 125\GeV, we define an $\mGG$ signal region (SR) around the Higgs boson mass (labeled $\mhsr$) using $110<\mGG<140\GeV$ and sideband (SB) regions below (labeled $\mhsblo$) and above (labeled $\mhsbhi$) the Higgs boson mass, by requiring $100<\mGG<110\GeV$ and $140<\mGG<180\GeV$, respectively. Since we assume that the two $\Pa$ particles from the $\haa$ decay have equal mass, we also divide the \mma distribution into an SR along the diagonal (labeled $\masr$), by requiring $\abs{\mao - \mat} < 0.3\GeV$, and a SB region (labeled $\masb$) using $\abs{\mao - \mat} > 0.3\GeV$. The final SR in which the MLE test is performed corresponds to the intersection of the above two SRs, $\mhsrmasr$. A binned likelihood fit of $\mu S+B$ versus the observed \mma distribution with bin sizes of $0.05 \times 0.05\GeV^{2}$ is used. The boundaries between the SR and SB regions are tuned using the simulation to maximize the signal enrichment in the SR, while maintaining a sufficient number of events in the SB regions to give a good background estimation. Figure~\ref{fig:2dma-bkg-obs} (center) shows the observed \mma distribution of the selected events in data from $\mhsr$ and the respective 1D projections of $\mao$ (left) and $\mat$ (right) in the $\masr$. Illustrative contours in \mma of the signal template (center) and its 1D projections (left and right) are also provided.

An estimate of the nonresonant background shape of the \mma template in
the $\mhsr$ is obtained from data SBs by taking the event-weighted
average of the shapes from the two $\mhsb$ regions (together labeled
$\mhsb$). To determine the normalization of the template, we assume that
the ratio of the number of events along the $\masr$ diagonal, from that
observed in the $\mhsr$s to that estimated from the $\mhsb$,
$N(\mhsrmasr)/N(\mhsbmasr)$, is the same as for the $\masb$ off diagonal, $N(\mhsrmasb)/N(\mhsbmasb)$. This is justified since the ratio of the number of events in the $\mhsr$ to the $\mhsb$ as a function of \mma is estimated to be constant after the reweighting procedure described below. The normalization for the template shape is then determined by solving for the expression for $N(\mhsrmasr)$.

For hadronic jets passing the event selection, there is an energy dependence, and thus an $\mGG$ dependence, that violates the assumption described in the previous paragraph. With increasing jet energy, more energy becomes available to produce additional hadrons in the jet, which increases the effective mass of the jet and thus its reconstructed $\mG$. To correct for this effect, prior to deriving the \mma shape, events from the two $\mhsb$ regions are first reweighted so that their transverse momentum distributions match that in the target $\mhsr$.

The normalized $\hgg$ and nonresonant background templates are then
added together. Their combined yield is renormalized so that the ratio
of the predicted to the observed yields in the $\masb$ region remains
unity. The $\hgg$ template accounts for about $0.4\%$ of the total
background template yield. To account for residual differences in the
\mma shape between the background estimate and the observed distribution
in the $\masb$ region, the total background template is multiplied by a
2D polynomial function $\pol$ over the full \mma range. A linear
polynomial is used with parameters chosen to maximize the likelihood
with respect to the observed \mma shape in the $\masb$ region. No
further improvement in the goodness of fit is found with higher polynomial orders. The resulting corrected \mma background template defines the background model used in the MLE fit. The left and right plots in Fig.~\ref{fig:2dma-bkg-obs} show that the SM background in the observed \mma distribution is dominated by single photonlike objects, which exhibit a smoothly falling $\mG$ spectrum, rather than by neutral-meson decays, which would be manifested as peaks~\cite{e2e-massreg}. Neutral-meson decays from $\gjet$ production are more likely to be reconstructed in the lower-energy $\Gamma_2$. The background estimation is validated using an orthogonal data sample obtained by inverting the $\chgiso$ requirement on $\Gamma_2$, to ensure negligible signal contamination. No significant bias is observed when the signal extraction procedure is performed on this sample.

Uncertainties in the predicted signal and background templates are
treated as nuisance parameters in the MLE procedure used to determine
the best fit signal strength $\muhat$. The dominant uncertainties
impacting the extracted $\muhat$ are those from statistical
uncertainties in the background template's shape. Their largest impact
on the relative uncertainty in $\muhat$ varies between 15\% and 20\%, depending on $\ma$. For $\ma = 0.1\GeV$, where the $\mG$ resolution is poorest and the background contribution is largest, systematic uncertainties affecting the background template normalization are also important. These include systematic uncertainties associated with the best fit parameters of $\pol$ and the relative contribution of $\mhsblo$ versus $\mhsbhi$ events in the non\-re\-so\-nant background \mma template. These systematic uncertainties impact the relative uncertainty in $\muhat$ by about 25\% (10\%) for $\ma$ = 0.1 (1.2)\GeV.

The systematic uncertainty in the signal strength from using the \mma template determined from simulation is estimated from a sample of electrons in events with $\zee$ decays, selected from both data and simulation with the ``tag-and-probe'' technique~\cite{tnp}. Electrons are preferred over decays of neutral mesons in jets because of the complicating effects of accompanying hadrons in the same jet~\cite{e2e-massreg}. Specifically, uncertainties associated with a relative $\mG$ scale shift and an increase in the smearing of the mass peak are estimated and found to have a negligible impact on $\muhat$.

The best fit background estimate determined from the MLE procedure is shown by the blue solid curves in Fig.~\ref{fig:2dma-bkg-obs} (left and right), together with its associated best fit total statistical plus systematic (stat + syst) uncertainties. We find no statistically significant excess in the data over the SM background predictions for $\ma$ masses in the range 0.1--1.2\GeV.

\begin{figure}[!htbp]
\centering
\includegraphics[width=\linewidth]{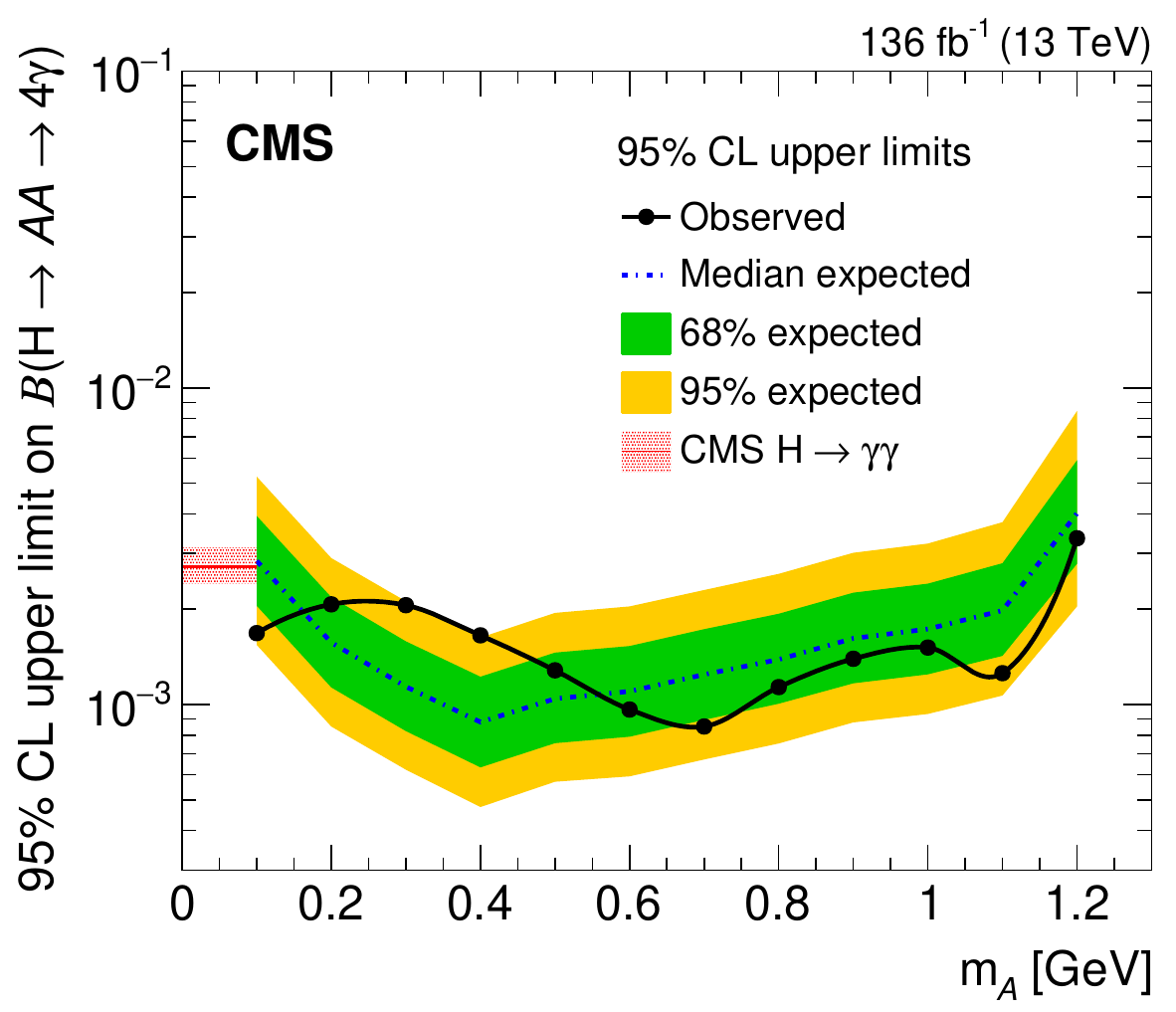}
\caption{Observed (black solid curve with points) and median expected (blue dashed curve) 95\% \CL upper limit on $\br(\hag)$ as a function of $\ma$ for prompt $\Pa$ decays. The 68\% (green band) and 95\% (yellow band) CIs are plotted around the expected limit. The 95\% \CL upper limit from the CMS measurement~\cite{hcoupling} of $\br(\hgg)$ is also shown (red band, where the width represents the uncertainty in the measurement).
}
\label{fig:limits}
\end{figure}

The \CLs criterion~\cite{CLS1,CLS2} is used to interpret this result in terms of excluded $\br(\hag)$ values. The observed upper limit on $\br(\hag)$ at 95\% confidence level (\CL) as a function of $\ma$ in the range 0.1-1.2\GeV is shown in Fig.~\ref{fig:limits}, and varies between (0.9--3.3)$ \times 10^{-3}$ for $\ma$ values 0.1--1.2\GeV. The expected $95\%$ \CL limits and their associated 68 and $95\%$ confidence intervals (CIs) are determined by simulating SM background-only pseudo-experiments. The LHC measurements of $\br(\hgg)$~\cite{hcoupling,atlascoupling} give an effective upper bound on a possible measurement of $\br(\hag)$ because of the degeneracy between the final states. The constraint from the CMS measurement~\cite{hcoupling} is shown in Fig.~\ref{fig:limits}. It is relevant for values of $\ma$ $\approx$ 0.1\GeV where the $\agg$ decay resembles a single photon and increases at larger $\ma$. Our observed upper limits thus set the best constraints for this decay mode in the $\ma$ range that we study.

We estimate the upper limits for long-lived $\Pa$ decays by comparing the signal yield in the $\mhsrmasr$ for different simulated $\Pa$ decay lengths compared with that for prompt decays. For $\ma$ = 0.1 (0.4)\GeV, the 95\% \CL upper limit on $\br(\hag)$ is 1.6 (0.9) times the prompt-decay upper limit for an $\Pa$ decay length of $\ctau$ = 1\mm, and 30 (3) times larger for $\ctau$ = 10\mm, with a linear interpolation between the two limits in both cases. Better upper limits result when the increased merging from long-lived $\Pa$ decays improves photon reconstruction more than it degrades $\mG$ resolution. The prompt-decay upper limits are also relevant for models with dissimilar $\Pa$ masses, $\PH \to \Pa_{1}\Pa_{2}$, with $m_{\Pa_1} \neq m_{\Pa_2}$, for mass differences less than the $\masr$ window, $|m_{\Pa_1} - m_{\Pa_2}| \lesssim 0.3\GeV$. For larger mass differences, the signal mass peak would fall outside of the $\masr$ and be absorbed into the $\masb$, making a measurement impossible.

In summary, the results of a search for the exotic Higgs boson decay
$\hag$ for a Higgs boson mass of 125\GeV are presented. Events
reconstructed with two photonlike objects are used, where each
photonlike object is assumed to be a merged $\agg$ candidate. A method
is developed to measure the invariant mass of merged diphoton candidates
to discriminate a potential signal from the standard model
background. This is the first search of its kind at CMS and made
possible by a novel end-to-end reconstruction technique of merged
diphotons. No excess of events above the estimated background is
found. An upper limit on the branching fraction $\br(\PH \to \Pa\Pa \to
4\gamma)$ of (0.9--3.3)$ \times 10^{-3}$ is set at 95\%~C.L.~for masses of $\Pa$ in the range 0.1--1.2\GeV, assuming prompt $\Pa$ decays. These are the current best constraints on $\hag$ in this $\ma$ range.

Tabulated results for this analysis are provided in the HEPData record~\cite{hepdata}.

\begin{acknowledgments}
We congratulate our colleagues in the CERN accelerator departments for the excellent performance of the LHC and thank the technical and administrative staffs at CERN and at other CMS institutes for their contributions to the success of the CMS effort. In addition, we gratefully acknowledge the computing centers and personnel of the Worldwide LHC Computing Grid and other centers for delivering so effectively the computing infrastructure essential to our analyses. Finally, we acknowledge the enduring support for the construction and operation of the LHC, the CMS detector, and the supporting computing infrastructure provided by the following funding agencies: BMBWF and FWF (Austria); FNRS and FWO (Belgium); CNPq, CAPES, FAPERJ, FAPERGS, and FAPESP (Brazil); MES and BNSF (Bulgaria); CERN; CAS, MoST, and NSFC (China); MINCIENCIAS (Colombia); MSES and CSF (Croatia); RIF (Cyprus); SENESCYT (Ecuador); MoER, ERC PUT and ERDF (Estonia); Academy of Finland, MEC, and HIP (Finland); CEA and CNRS/IN2P3 (France); BMBF, DFG, and HGF (Germany); GSRI (Greece); NKFIH (Hungary); DAE and DST (India); IPM (Iran); SFI (Ireland); INFN (Italy); MSIP and NRF (Republic of Korea); MES (Latvia); LAS (Lithuania); MOE and UM (Malaysia); BUAP, CINVESTAV, CONACYT, LNS, SEP, and UASLP-FAI (Mexico); MOS (Montenegro); MBIE (New Zealand); PAEC (Pakistan); MES and NSC (Poland); FCT (Portugal); MESTD (Serbia); MCIN/AEI and PCTI (Spain); MOSTR (Sri Lanka); Swiss Funding Agencies (Switzerland); MST (Taipei); MHESI and NSTDA (Thailand); TUBITAK and TENMAK (Turkey); NASU (Ukraine); STFC (United Kingdom); DOE and NSF (USA).
\end{acknowledgments}

\bibliography{auto_generated}

\providecommand{\href}[2]{#2}\begingroup\raggedright\begin{thebibliography}{10}%
\makeatletter
\providecommand{\hrefCMSnoop }[0]{\@secondoftwo}%
\makeatother
\providecommand{\doi}{\texttt{doi:}\begingroup \urlstyle{tt}\Url}

\bibitem{hcoupling}
\hrefCMSnoop {}{{CMS Collaboration}, ``Combined measurements of {H}iggs boson
  couplings in proton-proton collisions at $\sqrt{s}=13$ {TeV}'',} \textit{
  Eur. Phys. J. C} \textbf{ 79} (2019) 421,
  \href{http://dx.doi.org/10.1140/epjc/s10052-019-6909-y}{\doi{10.1140/epjc/s10052-019-6909-y}},
  \href{http://www.arXiv.org/abs/1809.10733}{\texttt{arXiv:1809.10733}}.

\bibitem{atlascoupling}
\hrefCMSnoop {}{{ATLAS Collaboration}, ``Combined measurements of {Higgs} boson
  production and decay using up to 80 $\mathrm{fb}^{-1}$ of proton-proton
  collision data at $\sqrt{s}=13$~{TeV} collected with the {ATLAS}
  experiment'',} \textit{ Phys. Rev. D} \textbf{ 101} (2020) 012002,
  \href{http://dx.doi.org/10.1103/physrevd.101.012002}{\doi{10.1103/physrevd.101.012002}},
  \href{http://www.arXiv.org/abs/1909.02845}{\texttt{arXiv:1909.02845}}.

\bibitem{hexo}
D.~Curtin\hrefCMSnoop {}{ {et~al.}, ``Exotic decays of the 125 {GeV} {H}iggs
  boson'',} \textit{ Phys. Rev. D} \textbf{ 90} (2014) 075004,
  \href{http://dx.doi.org/10.1103/PhysRevD.90.075004}{\doi{10.1103/PhysRevD.90.075004}},
  \href{http://www.arXiv.org/abs/1312.4992}{\texttt{arXiv:1312.4992}}.

\bibitem{hexo2}
\hrefCMSnoop {}{D.~Curtin, R.~Essig, S.~Gori, and J.~Shelton, ``Illuminating
  dark photons with high-energy colliders'',} \textit{ JHEP} \textbf{ 02}
  (2015) 157,
  \href{http://dx.doi.org/10.1007/jhep02(2015)157}{\doi{10.1007/jhep02(2015)157}},
  \href{http://www.arXiv.org/abs/1412.0018}{\texttt{arXiv:1412.0018}}.

\bibitem{pqtheory}
\hrefCMSnoop {}{R.~D. Peccei and H.~R. Quinn, ``$\mathrm{CP}$ conservation in
  the presence of pseudoparticles'',} \textit{ Phys. Rev. Lett.} \textbf{ 38}
  (1977) 1440,
  \href{http://dx.doi.org/10.1103/PhysRevLett.38.1440}{\doi{10.1103/PhysRevLett.38.1440}}.

\bibitem{axioncollider}
\hrefCMSnoop {}{M.~Bauer, M.~Neubert, and A.~Thamm, ``Collider probes of
  axion-like particles'',} \textit{ JHEP} \textbf{ 12} (2017) 044,
  \href{http://dx.doi.org/10.1007/jhep12(2017)044}{\doi{10.1007/jhep12(2017)044}},
  \href{http://www.arXiv.org/abs/1708.00443}{\texttt{arXiv:1708.00443}}.

\bibitem{axionreview}
\hrefCMSnoop {}{R.~D. Peccei, ``The strong {CP} problem and axions'',} in
  \textit{ Axions}, M.~Kuster, G.~Raffelt, and B.~Beltran, eds.
\newblock Springer Berlin Heidelberg, 2008.
\newblock
  \href{http://www.arXiv.org/abs/hep-ph/0607268}{\texttt{arXiv:hep-ph/0607268}}.
\newblock
  \href{http://dx.doi.org/10.1007/978-3-540-73518-2_1}{\doi{10.1007/978-3-540-73518-2_1}}.

\bibitem{cmshaa2m2t}
\hrefCMSnoop {}{{CMS Collaboration}, ``Search for an exotic decay of the
  {Higgs} boson to a pair of light pseudoscalars in the final state of two
  muons and two $\tau$ leptons in proton-proton collisions at $\sqrt{s}=13$
  {TeV}'',} \textit{ JHEP} \textbf{ 11} (2018) 018,
  \href{http://dx.doi.org/10.1007/jhep11(2018)018}{\doi{10.1007/jhep11(2018)018}},
  \href{http://www.arXiv.org/abs/1805.04865}{\texttt{arXiv:1805.04865}}.

\bibitem{cmshaa4m}
\hrefCMSnoop {}{{CMS Collaboration}, ``A search for pair production of new
  light bosons decaying into muons in proton-proton collisions at 13 {TeV}'',}
  \textit{ Phys. Lett. B} \textbf{ 796} (2019) 131,
  \href{http://dx.doi.org/10.1016/j.physletb.2019.07.013}{\doi{10.1016/j.physletb.2019.07.013}},
  \href{http://www.arXiv.org/abs/1812.00380}{\texttt{arXiv:1812.00380}}.

\bibitem{cmshaa2m2tlow}
\hrefCMSnoop {}{{CMS Collaboration}, ``Search for light pseudoscalar boson
  pairs produced from decays of the 125 {GeV} {Higgs} boson in final states
  with two muons and two nearby tracks in pp collisions at $\sqrt{s}=13$
  {TeV}'',} \textit{ Phys. Lett. B} \textbf{ 800} (2020) 135087,
  \href{http://dx.doi.org/10.1016/j.physletb.2019.135087}{\doi{10.1016/j.physletb.2019.135087}},
  \href{http://www.arXiv.org/abs/1907.07235}{\texttt{arXiv:1907.07235}}.

\bibitem{cmshaa2m2b}
\hrefCMSnoop {}{{CMS Collaboration}, ``Search for an exotic decay of the
  {H}iggs boson to a pair of light pseudoscalars in the final state with two
  muons and two b quarks in pp collisions at 13 {TeV}'',} \textit{ Phys. Lett.
  B} \textbf{ 795} (2019) 398,
  \href{http://dx.doi.org/10.1016/j.physletb.2019.06.021}{\doi{10.1016/j.physletb.2019.06.021}},
  \href{http://www.arXiv.org/abs/1812.06359}{\texttt{arXiv:1812.06359}}.

\bibitem{cmshaa2m2tboost}
\hrefCMSnoop {}{{CMS Collaboration}, ``Search for a light pseudoscalar {H}iggs
  boson in the boosted $\mu\mu\tau\tau$ final state in proton-proton collisions
  at $\sqrt{s}$ = 13 {TeV}'',} \textit{ JHEP} \textbf{ 08} (2020) 139,
  \href{http://dx.doi.org/10.1007/jhep08(2020)139}{\doi{10.1007/jhep08(2020)139}},
  \href{http://www.arXiv.org/abs/2005.08694}{\texttt{arXiv:2005.08694}}.

\bibitem{bogdanhaa}
\hrefCMSnoop {}{B.~A. Dobrescu, G.~Landsberg, and K.~T. Matchev, ``{Higgs}
  boson decays to $\mathrm{CP}$-odd scalars at the {Fermilab} {Tevatron} and
  beyond'',} \textit{ Phys. Rev. D} \textbf{ 63} (2001) 075003,
  \href{http://dx.doi.org/10.1103/PhysRevD.63.075003}{\doi{10.1103/PhysRevD.63.075003}},
  \href{http://www.arXiv.org/abs/hep-ph/0005308}{\texttt{arXiv:hep-ph/0005308}}.

\bibitem{axionsearch1}
P.~W. Graham\hrefCMSnoop {}{ {et~al.}, ``Experimental searches for the axion
  and axion-like particles'',} \textit{ Annu. Rev. Nucl. Part. Sci.} \textbf{
  65} (2015) 485,
  \href{http://dx.doi.org/10.1146/annurev-nucl-102014-022120}{\doi{10.1146/annurev-nucl-102014-022120}},
  \href{http://www.arXiv.org/abs/1602.00039}{\texttt{arXiv:1602.00039}}.

\bibitem{axionsearch2}
\hrefCMSnoop {}{I.~G. Irastorza and J.~Redondo, ``New experimental approaches
  in the search for axion-like particles'',} \textit{ Prog. Part. Nucl. Phys.}
  \textbf{ 102} (2018) 89,
  \href{http://dx.doi.org/10.1016/j.ppnp.2018.05.003}{\doi{10.1016/j.ppnp.2018.05.003}},
  \href{http://www.arXiv.org/abs/1801.08127}{\texttt{arXiv:1801.08127}}.

\bibitem{pdgreview2020}
\hrefCMSnoop {}{{Particle Data Group}, P.~A. Zyla {et~al.}, ``{Review of
  particle physics}'',} \textit{ PTEP} \textbf{ 2020} (2020) 083C01,
  \href{http://dx.doi.org/10.1093/ptep/ptaa104}{\doi{10.1093/ptep/ptaa104}}.

\bibitem{gluexagg}
\hrefCMSnoop {}{{GlueX} Collaboration, ``Search for photoproduction of
  axionlike particles at {GlueX}'',} \textit{ Phys. Rev. D} \textbf{ 105}
  (2022) 052007,
  \href{http://dx.doi.org/10.1103/physrevd.105.052007}{\doi{10.1103/physrevd.105.052007}},
  \href{http://www.arXiv.org/abs/2109.13439}{\texttt{arXiv:2109.13439}}.

\bibitem{axionlead}
\hrefCMSnoop {}{S.~Knapen, T.~Lin, H.~K. Lou, and T.~Melia, ``Searching for
  axionlike particles with ultraperipheral heavy-ion collisions'',} \textit{
  Phys. Rev. Lett.} \textbf{ 118} (2017) 171801,
  \href{http://dx.doi.org/10.1103/PhysRevLett.118.171801}{\doi{10.1103/PhysRevLett.118.171801}},
  \href{http://www.arXiv.org/abs/1607.06083}{\texttt{arXiv:1607.06083}}.

\bibitem{axionastro}
\hrefCMSnoop {}{G.~G. Raffelt, ``Astrophysical axion bounds'',} in \textit{
  Axions}, M.~Kuster, G.~Raffelt, and B.~Beltran, eds.
\newblock Springer Berlin Heidelberg, 2008.
\newblock
  \href{http://www.arXiv.org/abs/hep-ph/0611350}{\texttt{arXiv:hep-ph/0611350}}.
\newblock
  \href{http://dx.doi.org/10.1007/978-3-540-73518-2_3}{\doi{10.1007/978-3-540-73518-2_3}}.

\bibitem{axioncosmo}
\hrefCMSnoop {}{P.~Sikivie, ``Axion cosmology'',} in \textit{ Axions},
  M.~Kuster, G.~Raffelt, and B.~Beltran, eds.
\newblock Springer Berlin Heidelberg, 2008.
\newblock
  \href{http://www.arXiv.org/abs/astro-ph/0610440}{\texttt{arXiv:astro-ph/0610440}}.
\newblock
  \href{http://dx.doi.org/10.1007/978-3-540-73518-2_2}{\doi{10.1007/978-3-540-73518-2_2}}.

\bibitem{axioncosmo2}
\hrefCMSnoop {}{D.~J.~E. Marsh, ``Axion cosmology'',} \textit{ Phys. Rep.}
  \textbf{ 643} (2016) 1,
  \href{http://dx.doi.org/10.1016/j.physrep.2016.06.005}{\doi{10.1016/j.physrep.2016.06.005}},
  \href{http://www.arXiv.org/abs/1510.07633}{\texttt{arXiv:1510.07633}}.

\bibitem{axiondmreview}
\hrefCMSnoop {}{F.~Chadha-Day, J.~Ellis, and D.~J.~E. Marsh, ``Axion dark
  matter: What is it and why now?'',} \textit{ Science Advances} \textbf{ 8}
  (2022) eabj3618,
  \href{http://dx.doi.org/10.1126/sciadv.abj3618}{\doi{10.1126/sciadv.abj3618}},
  \href{http://www.arXiv.org/abs/2105.01406}{\texttt{arXiv:2105.01406}}.

\bibitem{e2e-massreg}
\hrefCMSnoop {}{{CMS Collaboration}, ``{Reconstruction of decays to merged
  photons using end-to-end deep learning with domain continuation in the CMS
  detector}'',} 2022.
  \href{http://www.arXiv.org/abs/2204.12313}{\texttt{arXiv:2204.12313}}.

\bibitem{hgg}
\hrefCMSnoop {}{{CMS Collaboration}, ``Observation of the diphoton decay of the
  {H}iggs boson and measurement of its properties'',} \textit{ Eur. Phys. J. C}
  \textbf{ 74} (2014) 3076,
  \href{http://dx.doi.org/10.1140/epjc/s10052-014-3076-z}{\doi{10.1140/epjc/s10052-014-3076-z}},
  \href{http://www.arXiv.org/abs/1407.0558}{\texttt{arXiv:1407.0558}}.

\bibitem{hepdata}
\hrefCMSnoop {}{}{HEPD}ata record for this analysis, 2022.
\newblock
  \href{http://dx.doi.org/10.17182/hepdata.132767}{\doi{10.17182/hepdata.132767}}.

\bibitem{cms}
\hrefCMSnoop {}{{CMS Collaboration}, ``{The CMS experiment at the CERN LHC}'',}
  \textit{ JINST} \textbf{ 3} (2008) S08004,
  \href{http://dx.doi.org/10.1088/1748-0221/3/08/S08004}{\doi{10.1088/1748-0221/3/08/S08004}}.

\bibitem{CMS:2020cmk}
\hrefCMSnoop {}{{CMS Collaboration}, ``{Performance of the {CMS} {Level-1}
  trigger in proton-proton collisions at $\sqrt{s} = 13$\,TeV}'',} \textit{
  JINST} \textbf{ 15} (2020) P10017,
  \href{http://dx.doi.org/10.1088/1748-0221/15/10/P10017}{\doi{10.1088/1748-0221/15/10/P10017}},
  \href{http://www.arXiv.org/abs/2006.10165}{\texttt{arXiv:2006.10165}}.

\bibitem{cmstrigger}
\hrefCMSnoop {}{{CMS Collaboration}, ``{The {CMS} trigger system}'',} \textit{
  JINST} \textbf{ 12} (2016) P01020,
  \href{http://dx.doi.org/10.1088/1748-0221/12/01/P01020}{\doi{10.1088/1748-0221/12/01/P01020}},
  \href{http://www.arXiv.org/abs/1609.02366}{\texttt{arXiv:1609.02366}}.

\bibitem{CMS:2015xaf}
\hrefCMSnoop {}{{CMS Collaboration}, ``{Performance of electron reconstruction
  and selection with the {CMS} detector in proton-proton collisions at
  $\sqrt{s} = 8$\TeV}'',} \textit{ JINST} \textbf{ 10} (2015) P06005,
  \href{http://dx.doi.org/10.1088/1748-0221/10/06/P06005}{\doi{10.1088/1748-0221/10/06/P06005}},
\href{http://www.arXiv.org/abs/1502.02701}{\texttt{arXiv:1502.02701}}.

\bibitem{CMS:2018rym}
\hrefCMSnoop {}{{CMS Collaboration}, ``{Performance of the CMS muon detector
  and muon reconstruction with proton-proton collisions at $\sqrt{s}=$ 13
  {TeV}}'',} \textit{ JINST} \textbf{ 13} (2018) P06015,
  \href{http://dx.doi.org/10.1088/1748-0221/13/06/P06015}{\doi{10.1088/1748-0221/13/06/P06015}},
  \href{http://www.arXiv.org/abs/1804.04528}{\texttt{arXiv:1804.04528}}.

\bibitem{CMS:2015myp}
\hrefCMSnoop {}{{CMS Collaboration}, ``Performance of photon reconstruction and
  identification with the {CMS} detector in proton-proton collisions at
  $\sqrt{s}$ = 8~{TeV}'',} \textit{ JINST} \textbf{ 10} (2015) P08010,
  \href{http://dx.doi.org/10.1088/1748-0221/10/08/P08010}{\doi{10.1088/1748-0221/10/08/P08010}},
  \href{http://www.arXiv.org/abs/1502.02702}{\texttt{arXiv:1502.02702}}.

\bibitem{CMS:2014pgm}
\hrefCMSnoop {}{{CMS Collaboration}, ``{Description and performance of track
  and primary-vertex reconstruction with the {CMS} tracker}'',} \textit{ JINST}
  \textbf{ 9} (2014) P10009,
  \href{http://dx.doi.org/10.1088/1748-0221/9/10/P10009}{\doi{10.1088/1748-0221/9/10/P10009}},
  \href{http://www.arXiv.org/abs/1405.6569}{\texttt{arXiv:1405.6569}}.

\bibitem{CMS:2020uim}
\hrefCMSnoop {}{{CMS Collaboration}, ``Electron and photon reconstruction and
  identification with the {CMS} experiment at the {CERN} {LHC}'',} \textit{
  JINST} \textbf{ 16} (2021) P05014,
  \href{http://dx.doi.org/10.1088/1748-0221/16/05/P05014}{\doi{10.1088/1748-0221/16/05/P05014}},
  \href{http://www.arXiv.org/abs/2012.06888}{\texttt{arXiv:2012.06888}}.

\bibitem{CMS:2017yfk}
\hrefCMSnoop {}{{CMS Collaboration}, ``Particle-flow reconstruction and global
  event description with the {CMS} detector'',} \textit{ JINST} \textbf{ 12}
  (2017) P10003,
  \href{http://dx.doi.org/10.1088/1748-0221/12/10/P10003}{\doi{10.1088/1748-0221/12/10/P10003}},
\href{http://www.arXiv.org/abs/1706.04965}{\texttt{arXiv:1706.04965}}.

\bibitem{CMS:2018jrd}
\hrefCMSnoop {}{{CMS Collaboration}, ``{Performance of reconstruction and
  identification of $\tau$ leptons decaying to hadrons and $\nu_\tau$ in pp
  collisions at $\sqrt{s}=$ 13~{TeV}}'',} \textit{ JINST} \textbf{ 13} (2018)
  P10005,
  \href{http://dx.doi.org/10.1088/1748-0221/13/10/P10005}{\doi{10.1088/1748-0221/13/10/P10005}},
  \href{http://www.arXiv.org/abs/1809.02816}{\texttt{arXiv:1809.02816}}.

\bibitem{CMS:2016lmd}
\hrefCMSnoop {}{{CMS Collaboration}, ``{Jet energy scale and resolution in the
  {CMS} experiment in pp collisions at 8 {TeV}}'',} \textit{ JINST} \textbf{
  12} (2017) P02014,
  \href{http://dx.doi.org/10.1088/1748-0221/12/02/P02014}{\doi{10.1088/1748-0221/12/02/P02014}},
\href{http://www.arXiv.org/abs/1607.03663}{\texttt{arXiv:1607.03663}}.

\bibitem{CMS:2019ctu}
\hrefCMSnoop {}{{CMS Collaboration}, ``Performance of missing transverse
  momentum reconstruction in proton-proton collisions at $\sqrt{s} = 13$\,{TeV}
  using the {CMS} detector'',} \textit{ JINST} \textbf{ 14} (2019) P07004,
  \href{http://dx.doi.org/10.1088/1748-0221/14/07/P07004}{\doi{10.1088/1748-0221/14/07/P07004}},
\href{http://www.arXiv.org/abs/1903.06078}{\texttt{arXiv:1903.06078}}.

\bibitem{CMS-LUM-17-003}
\hrefCMSnoop {}{{CMS Collaboration}, ``Precision luminosity measurement in
  proton-proton collisions at $\sqrt{s} =$ 13 {TeV} in 2015 and 2016 at
  {CMS}'',} \textit{ Eur. Phys. J. C} \textbf{ 81} (2021) 800,
  \href{http://dx.doi.org/10.1140/epjc/s10052-021-09538-2}{\doi{10.1140/epjc/s10052-021-09538-2}},
  \href{http://www.arXiv.org/abs/2104.01927}{\texttt{arXiv:2104.01927}}.

\bibitem{CMS-PAS-LUM-17-004}
\href {https://cds.cern.ch/record/2621960/}{{CMS Collaboration}, ``{CMS}
  luminosity measurement for the 2017 data-taking period at $\sqrt{s}$ = 13
  {TeV}'',} CMS Physics Analysis Summary CMS-PAS-LUM-17-004, 2018.

\bibitem{CMS-PAS-LUM-18-002}
\href {https://cds.cern.ch/record/2676164/}{{CMS Collaboration}, ``{CMS}
  luminosity measurement for the 2018 data-taking period at $\sqrt{s}$ = 13
  {TeV}'',} CMS Physics Analysis Summary CMS-PAS-LUM-18-002, 2019.

\bibitem{h4gTanvi}
\href {http://cds.cern.ch/record/2776775}{{CMS Collaboration}, ``{Search for
  the exotic decay of the {Higgs} boson into two light pseudoscalars with four
  photons in the final state in proton-proton collisions at $\sqrt{s}$ = 13
  TeV}'',} 2022.
  \href{http://www.arXiv.org/abs/2208.01469}{\texttt{arXiv:2208.01469}}.

\bibitem{lhoodlhc}
\href {https://cds.cern.ch/record/1379837}{{ATLAS and CMS Collaborations, and
  LHC Higgs Combination Group}, ``Procedure for the {LHC} {Higgs} boson search
  combination in {Summer} 2011'',} Technical Report CMS-NOTE-2011-005,
  ATL-PHYS-PUB-2011-11, 2011.

\bibitem{mg5mcatnlo}
J.~Alwall\hrefCMSnoop {}{ {et~al.}, ``The automated computation of tree-level
  and next-to-leading order differential cross sections, and their matching to
  parton shower simulations'',} \textit{ JHEP} \textbf{ 07} (2014) 079,
  \href{http://dx.doi.org/10.1007/JHEP07(2014)079}{\doi{10.1007/JHEP07(2014)079}},
  \href{http://www.arXiv.org/abs/1405.0301}{\texttt{arXiv:1405.0301}}.

\bibitem{fxfx}
\hrefCMSnoop {}{R.~Frederix and S.~Frixione, ``Merging meets matching in
  {MC}@{NLO}'',} \textit{ JHEP} \textbf{ 12} (2012) 061,
  \href{http://dx.doi.org/10.1007/jhep12(2012)061}{\doi{10.1007/jhep12(2012)061}},
  \href{http://www.arXiv.org/abs/1209.6215}{\texttt{arXiv:1209.6215}}.

\bibitem{higgswg}
{LHC Higgs Cross Section Working Group} Collaboration, D.~de~Florian {et~al.},
  ``{Handbook of {LHC Higgs} cross sections: 4. Deciphering the nature of the
  {Higgs} sector}''.
\newblock CERN Yellow Reports: Monographs. CERN, 2016.
\newblock
  \href{http://dx.doi.org/10.23731/CYRM-2017-002}{\doi{10.23731/CYRM-2017-002}}.

\bibitem{tnp}
\hrefCMSnoop {}{{CMS Collaboration}, ``Measurement of the inclusive
  $\mathrm{W}$ and $\mathrm{Z}$ production cross sections in pp collisions at
  $\sqrt{s}=7$ {TeV}'',} \textit{ JHEP} \textbf{ 10} (2011) 132,
  \href{http://dx.doi.org/10.1007/JHEP10(2011)132}{\doi{10.1007/JHEP10(2011)132}},
  \href{http://www.arXiv.org/abs/1107.4789}{\texttt{arXiv:1107.4789}}.

\bibitem{CLS1}
\hrefCMSnoop {}{A.~L. Read, ``Presentation of search results: The
  {CL$_{\text{s}}$} technique'',} \textit{ J. Phys. G} \textbf{ 28} (2002)
  2693,
\href{http://dx.doi.org/10.1088/0954-3899/28/10/313}{\doi{10.1088/0954-3899/28/10/313}}.

\bibitem{CLS2}
\hrefCMSnoop {}{T.~Junk, ``Confidence level computation for combining searches
  with small statistics'',} \textit{ Nucl. Instrum. Meth. A} \textbf{ 434}
  (1999) 435,
  \href{http://dx.doi.org/10.1016/S0168-9002(99)00498-2}{\doi{10.1016/S0168-9002(99)00498-2}},
\href{http://www.arXiv.org/abs/hep-ex/9902006}{\texttt{arXiv:hep-ex/9902006}}.

\end{thebibliography}\endgroup
\cleardoublepage \appendix\section{The CMS Collaboration \label{app:collab}}\begin{sloppypar}\hyphenpenalty=5000\widowpenalty=500\clubpenalty=5000
\cmsinstitute{Yerevan Physics Institute, Yerevan, Armenia}
{\tolerance=6000
A.~Tumasyan\cmsAuthorMark{1}\cmsorcid{0009-0000-0684-6742}
\par}
\cmsinstitute{Institut f\"{u}r Hochenergiephysik, Vienna, Austria}
{\tolerance=6000
W.~Adam\cmsorcid{0000-0001-9099-4341}, J.W.~Andrejkovic, T.~Bergauer\cmsorcid{0000-0002-5786-0293}, S.~Chatterjee\cmsorcid{0000-0003-2660-0349}, K.~Damanakis\cmsorcid{0000-0001-5389-2872}, M.~Dragicevic\cmsorcid{0000-0003-1967-6783}, A.~Escalante~Del~Valle\cmsorcid{0000-0002-9702-6359}, P.S.~Hussain\cmsorcid{0000-0002-4825-5278}, M.~Jeitler\cmsAuthorMark{2}\cmsorcid{0000-0002-5141-9560}, N.~Krammer\cmsorcid{0000-0002-0548-0985}, L.~Lechner\cmsorcid{0000-0002-3065-1141}, D.~Liko\cmsorcid{0000-0002-3380-473X}, I.~Mikulec\cmsorcid{0000-0003-0385-2746}, P.~Paulitsch, F.M.~Pitters, J.~Schieck\cmsAuthorMark{2}\cmsorcid{0000-0002-1058-8093}, R.~Sch\"{o}fbeck\cmsorcid{0000-0002-2332-8784}, D.~Schwarz\cmsorcid{0000-0002-3821-7331}, S.~Templ\cmsorcid{0000-0003-3137-5692}, W.~Waltenberger\cmsorcid{0000-0002-6215-7228}, C.-E.~Wulz\cmsAuthorMark{2}\cmsorcid{0000-0001-9226-5812}
% ...
\par}
\vskip\cmsinstskip
\dag:~Deceased\\
$^{1}$Also at Yerevan State University, Yerevan, Armenia\\
$^{2}$Also at TU Wien, Vienna, Austria\\
% ...
\end{sloppypar}
\end{document}